\documentclass{article}
\usepackage{braket}
\usepackage{amsmath}

\begin{document}

\title{Zero-Correlation Entanglement}


\author{Toru Ohira\thanks{The author is also affiliated with Future Value Creation Research Center, Graduate School of Informatics, Nagoya University, and with Mathematical Science Team, RIKEN Center for Advanced Intelligence Project.}\\
Graduate School of Mathematics, Nagoya University, Nagoya, Japan} 

\maketitle

\begin{abstract}%
We consider a quantum entangled state for two particles, each particle having two basis states, which includes an entangled pair of spin 1/2 particles. 
We show that, for any quantum entangled state vectors of such systems, one can always find a 
pair of observable operators $\mathcal{X}, \mathcal{Y}$ with zero correlations ($\bra{\psi}\mathcal{X}\mathcal{Y}\ket{\psi} - \bra{\psi}\mathcal{X}\ket{\psi}\bra{\psi}\mathcal{Y}\ket{\psi}= 0$).
At the same time, if we consider the analogous classical system of a ``classically entangled'' (statistically non-independent) pair of random variables taking two values, one 
can never have zero correlations (zero covariance, $E[XY] - E[X]E[Y] = 0$). We provide 
a general proof to illustrate the different nature of entanglements in classical and quantum theories.
\end{abstract}

\section{Introduction}

Entanglement is considered a key concept in understanding quantum phenomena. In understanding entanglement, correlations based on expectation values of quantum observable operators for multi-particle systems have been investigated both theoretically and experimentally (e.g.,\cite{bell,clauser,aspect,leggett,vedral,altepeter,sakai,horodecki}). For example, among the research efforts for obtaining conditions for the separability of density matrices(e.g.,\cite{werner,gisin,peres,fujikawa1,fujikawa2}), a recent work by Fujikawa et. al.\cite{fujikawa2} has associated saparability with zero correlations and analyzed experimental results.

In this paper, we also study correlations to investigate entangled quantum systems. Our approach, however, differs from previous works that associated zero correlations with separability. Rather, entanglements are connected with zero correlations. 
In particular, we consider a system of two quantum particles, each taking two distinct states ($2\times 2$ system), which includes systems such as a pair of spin 1/2 particles. 
Our main result on 
this system is a theorem stating that, for any entangled state vector, one can 
always find a pair of quantum observable operators with a zero correlation.

As well as providing a general proof on the quantum system, we also note that for analogous classical systems consisting of two stochastic variables, each taking on one of two possible values, there cannot be zero correlation unless the variables are statistically independent. We may regard the 
statistical dependence of classical variables as analogous to quantum entanglement. Thus, zero correlation in the latter case but not the former provides yet another illustration of the difference between classical and
quantum probability theories.

\section{Definitions and System Descriptions}

\subsection{Classical case}

We consider two random variables $X$ and $Y$, such that they both take only two distinct finite values $(x_1, x_2)$ and $(y_1,y_2)$. The joint probability distribution for these variables is denoted as
$P(X:Y)$, and given by
\begin{equation}
P(X={x_i}:Y={y_j}) = p_{ij}, \quad ( i,j \in \{1,2\} )\nonumber
\label{jp}
\end{equation}
The probability distributions $P(X)$ for $X$ and $P(Y)$ for $Y$ can
be derived from above as.
\begin{equation}
P(X={x_i}) = p_{i1} + p_{i2}, \quad
P(Y={y_j}) = p_{1j} + p_{2j}.\nonumber
\label{ep}
\end{equation}
By the requirement that both $X$, and  $Y$ take only two values,
\begin{equation}
p({x_1}) + p({x_2}) = p({y_1}) + p({y_2}) = 1.\nonumber
\label{sp}
\end{equation}

When the joint probability of $X$ and $Y$ can be decomposed as
\begin{equation}
P(X:Y)=P(X)P(Y),\nonumber
\label{ind0}
\end{equation}
these stochastic variables are called statistically independent.
If such decomposition is not possible, $X$ and $Y$ are not statistically independent (``classically entangled'').

Expectation values are defined as 
\begin{equation}
E[X] = \sum_i p({x_i}){x_i},\quad
E[Y] = \sum_i p({y_i}){y_i},\quad
E[XY] = \sum_{i,j} p({x_i}:{y_j}){x_i}{y_j}.
\label{ev}
\nonumber
\end{equation}

From these we define the covariance of $X$ and $Y$ as
\begin{equation}
Cov[X,Y] \equiv E[XY] - E[X]E[Y].
\nonumber
\end{equation}
We define zero correlation to be when this covariance is zero
or equivalently, 
\begin{equation}
E[XY] = E[X]E[Y].\nonumber
\label{noc0}
\end{equation}

We later mention in the main theorem that for this classical $2\times 2$ sytem, statistical independence and zero correlation is equivalent. 
\vspace{1em}

\subsection{Quantum case}

We consider two quantum particles $A$ and $B$. 
The total normalized state vector $\ket{\psi}$ is given by

\begin{equation}
\ket{\psi}= \sum_{i,j} \omega_{i,j}\ket{a_i}\otimes\ket{b_j}, \nonumber
\end{equation}
where $\ket{a}$ and $\ket{b}$ describe the state of particle $A$ and $B$ respectively, and $\omega_{i,j}$ are quantum amplitudes given by complex scalars. This system is called a separable (product) state
when the total state vector can be decomposed as a product of each
normalized state of $A$ and $B$,
\begin{equation}
\ket{\psi} = \ket{\psi_A}\otimes\ket{\psi_B} \nonumber
\end{equation}
where
\begin{equation}
\ket{\psi_A}= \sum_{i} \nu_i \ket{a_i}  \quad  \ket{\psi_B}= \sum_{j} \tau_j \ket{b_j}.\nonumber
\end{equation}

There is a correspondence between the classical notion of statistical independence and this separability of state vectors. State vectors which are not separable are called (quantum) entangled state vectors. Thus, quantum entanglement corresponds to ``classical entanglement," i.e. lack of statical independence of two stochastic variables, as presented in the previous subsection.

We now present the expectation values for the quantum system.
For this, two operators $\mathcal{X}$ and $\mathcal{Y}$ are defined as
$\mathcal{X} = {\mathcal{Q}_A}\otimes{\bf{1}}_B$ and $\mathcal{Y} = {\bf{1}}_A\otimes\mathcal{R}_B$.
Here, 
${\mathcal{Q}_A}$ and ${\mathcal{R}_B}$ are quantum observable operators for $A$ and $B$ respectively, and ${\bf{1}}$ is the identity operator.

With these operators, we consider the relations between the expectation values of $\bra{\psi}\mathcal{X}\mathcal{Y}\ket{\psi}$ and $\bra{\psi}\mathcal{X}\ket{\psi}\bra{\psi}\mathcal{Y}\ket{\psi}$. 
For separable states in general, it is straightforward to show that they are always equal, that is, zero correlated. 

We ask the same question for the case of entangled states by limiting ourselves to 
the case that particles $A$ and $B$ take only two distinctive states. In this case
the most general quantum state vector is given as follows, with $(i)^2 = -1$:
\begin{equation}
\ket{\psi}= \alpha\ket{a_1}\otimes\ket{b_1} + \beta {e^{i\phi}} \ket{a_1}\otimes\ket{b_2} 
+ \gamma {e^{i\kappa}} \ket{a_2}\otimes\ket{b_1}+ \delta {e^{i\lambda}} \ket{a_2}\otimes\ket{b_2}
\label{gstate}
\end{equation}
where $\alpha,\beta,\gamma,\delta, \phi,\kappa$, and $\lambda$ are real-valued parameters with 
$\alpha^2 + \beta^2 + \gamma^2 + \delta^2 = 1$ and $0 \leq \phi,\kappa,\lambda \leq 2\pi$.

We note that this state vector is separable when $\alpha\delta = \beta\gamma$ and $\lambda = (\phi + \kappa) \bmod 2\pi$, but is quantum entangled otherwise.

Also, the most general observable operators are given by Hermitian matrices in the basis
of $\ket{a_{(1,2)}}$ and $\ket{b_{(1,2)}}$ respectively,
\begin{equation}
{\mathcal{Q}_A}=
\begin{bmatrix}
{Q_+} & q{e^{i s}} \\
q{e^{-i s}} & {Q_-}
\end{bmatrix}
,\quad
{\mathcal{R}_B}=
\begin{bmatrix}
{R_+} & r{e^{i v}} \\
r{e^{-i v}} & {R_-}
\end{bmatrix}
\end{equation}
where $Q_\pm, R_\pm, q, r, s$, and $v$ are real-valued parameters with $0 \leq s,v \leq 2\pi$.

With the above setup, we will show in the next section that even for any entangled state
vector in the form above, we can always find a pair of ${\mathcal{Q}_A}, {\mathcal{R}_B}$
that achieves zero correlation, 
$\bra{\psi}\mathcal{X}\mathcal{Y}\ket{\psi} = \bra{\psi}\mathcal{X}\ket{\psi}\bra{\psi}\mathcal{Y}\ket{\psi}$ with $\mathcal{X} = {\mathcal{Q}_A}\otimes{\bf{1}}_B$ and $\mathcal{Y} = {\bf{1}}_A\otimes\mathcal{R}_B$.

\section{Main Theorem and Proof}

With the set up in the previous section for a dual particle system, we discuss the relation between entanglements and zero-correlations. For non-entangled systems, both classical (statistically independent) and quantum (separable) cases commonly lead to zero-correlations. 
For entangled systems, however, there is a clear difference between the classical and quantum systems, which is reiterated in the following theorem.
\vspace{1em}

\noindent
{\bf{Theorem}}
\vspace{1em}

\noindent
Classical Case: 

For any pair of random variables $X$ and $Y$ each taking two distinct finite values
(for any values of $(p_{ij}, x_i, y_j)$, $i, j \in \{1,2\}$ as set up above)
that are not statistically independent (``classically entangled''), 
they can NEVER be zero correlated.
\vspace{1em}

\noindent
Quantum Case:

For any quantum-entangled pure state for a dual particle system each taking two 
distinct states as set up above, one can ALWAYS find a pair of observable operators
 $\mathcal{X} = {\mathcal{Q}_A}\otimes{\bf{1}}_B$ and $\mathcal{Y} = {\bf{1}}_A\otimes\mathcal{R}_B$ that are zero correlated.
\vspace{2em}

\noindent
{\bf{Proof}}
\vspace{1em}

\noindent
Classical Case: 

We have established the equivalence of statistical independence and
zero correlation for such statistical variables $X$ and $Y$ in the previous work\cite{ohira}. 
The statement here follows immediately. (In passing, we note that the classical $2\times2$ system is special and one can easily create examples showing that this equivalence does not hold in higher dimensions\cite{feller,bain}.)
\vspace{1em}

\noindent
Quantum Case:

We want to show that for any general dual particle state vector $\ket{\psi}$ in Eq. (\ref{gstate}),
we can always find a pair of observable operators $\mathcal{X},\mathcal{Y}$ such that the following zero-correlation relation holds.
\begin{equation}
\bra{\psi}\mathcal{X}\mathcal{Y}\ket{\psi} = \bra{\psi}\mathcal{X}\ket{\psi}\bra{\psi}\mathcal{Y}\ket{\psi}
\label{zce1}
\end{equation}
with $\mathcal{X} = {\mathcal{Q}_A}\otimes{\bf{1}}_B$ and $\mathcal{Y} = {\bf{1}}_A\otimes\mathcal{R}_B$.
\vspace{1em}

After tedious calculations, the above statement translates to the following:
\vspace{1em}

Given any set of real-valued parameters as in Eq. (\ref{gstate}) -- $\alpha,\beta,\gamma,\delta, \phi,\kappa,\lambda$ with $\alpha^2 + \beta^2 + \gamma^2 + \delta^2 = 1$ and $0 \leq \phi,\kappa,\lambda \leq 2\pi$ -- one can find a set of real-valued parameters $Q_\pm, R_\pm, q, r, s, v$ with $0 \leq s,v \leq 2\pi$
such that the following holds.
\begin{eqnarray}
&\ &{\alpha^2}{Q_{+}}{R_{+}} + {\beta^2}{{Q_+}{R_-}} + {\gamma^2}{Q_-}{R_+} + {\delta^2}{Q_-}{R_-} \nonumber\\
&+ & 2\alpha\beta\cos(\phi + v){Q_{+}}{r} + 2\alpha\gamma\cos(\kappa + s){q}{R_{+}}\nonumber\\
&+ & 2\beta\delta\cos(\lambda - \phi + s){q}{R_{-}}+ 2\gamma\delta\cos(\lambda - \kappa + v){Q_-}{r} \nonumber\\
&+ & 2\alpha\delta\cos(\lambda + s + v){q}{r}+ 2\beta\gamma\cos(\kappa - \phi + s - v){q}{r} \nonumber\\
&= & [({\alpha^2}+{\beta^2}){Q_{+}} + ({\gamma^2}+{\delta^2}){Q_{-}} + 2\alpha\gamma\cos(\kappa + s){q} + 2\beta\delta\cos(\lambda - \phi + s){q}]\nonumber\\
&\times & [({\alpha^2}+{\gamma^2}){R_{+}} +({\beta^2}+{\delta^2}){R_{-}} + 2\alpha\beta\cos(\phi + v){r} + 2\gamma\delta\cos(\lambda - \kappa + v){r}]
\nonumber\\
&\ &
\label{zce2}
\end{eqnarray}

Finding the general solution, i.e., all possible sets of parameters $Q_\pm, R_\pm, q, r, s, v$, is difficult. We can, however, find a set of parameters for which the above statement holds. For this we first set the phase parameters as
\begin{equation}
s = {1\over 2}(- \kappa + \phi - \lambda), \quad v = {1\over 2}(\kappa - \phi - \lambda)
\end{equation}
Further, if we define $\xi = \cos({1\over 2}(\lambda - \phi - \kappa))$, $Q_{\pm} = Q_0 \pm \epsilon$ and $R_{\pm} = R_0 \pm \eta$, 
(\ref{zce2}) can be simplified as follows:
\begin{eqnarray}
&\ & 2(\alpha\delta - \beta\gamma)(\alpha\delta + \beta\gamma)\epsilon\eta + (\alpha\delta + \beta\gamma)qr \nonumber\\
&= & 2(\alpha\delta - \beta\gamma)(\alpha\beta - \gamma\delta)\xi q \eta 
+ 2(\alpha\delta - \beta\gamma)(\alpha\gamma - \beta\delta)\xi r \epsilon \nonumber\\
&+ & 2(\alpha\gamma + \beta\delta)(\alpha\beta + \gamma\delta){\xi^2} qr \label{zce3}
\end{eqnarray}
(Note, $Q_{0}, R_{0}, s, v$ do not appear in Eq. (\ref{zce3}).)
\vspace{1em}

Our aim now is to find $\epsilon, \eta, q$, and $r$ to satisfy Eq. (\ref{zce3}) given any set of $\alpha, \beta, \gamma, \delta, \xi$ with $\alpha^2 + \beta^2 + \gamma^2 + \delta^2 = 1$ and $-1 \leq \xi \leq 1$. Let us also impose the conditions $\epsilon^2 + q^2 \neq 0$, and $\eta^2 + r^2 \neq 0$, so that the observable operators have two distinct eigenvalues.

We first note that for the separable case ($\alpha\delta = \beta\gamma$ and $\xi = 1$) Eq. (\ref{zce3}) holds for any set of $\epsilon, \eta, q, r$ as
expected by the fact that the separability of the state vector $\ket{\psi}$ entails the zero correlation.

Even for the entangled case, one can find the desired set of parameters by explicit constructions. We do this by considering different cases which, taken altogether, comprise all possible values of $\alpha, \beta, \gamma, \delta$, and $\xi$.

\subsection{\underline{$\alpha\delta - \beta\gamma \neq 0$ and $\xi \neq 0$}}
\vspace{1em}

\subsubsection{$\alpha\delta + \beta\gamma = 0$} \
\vspace{-1em}

 $(\epsilon, \eta, q, r) = (\epsilon, \eta, q = 0, r=0)$.

No constraints on $\epsilon, \eta$ other than $\epsilon^2 + q^2 \neq 0$, and $\eta^2 + r^2 \neq 0$. (Hereafter, the same convention is used: unless specified, no constraints other than $\epsilon^2 + q^2 \neq 0$, and $\eta^2 + r^2 \neq 0$.)

\subsubsection{$\alpha\delta + \beta\gamma \neq 0$} \
\vspace{-1em}

This case needs to be considered by further classifications.
\vspace{1em}

\noindent
(i) $\alpha\beta - \gamma\delta \neq 0$ and $\alpha\gamma - \beta\delta \neq 0$

We can take either of the following two parameter settings.
\begin{itemize}
\item $(\epsilon, \eta, q, r) = (\epsilon \neq 0, \eta\neq 0, q \neq 0, r=0)$ such that
$(\alpha\delta + \beta\gamma){\epsilon} = {q} \xi (\alpha\beta - \gamma\delta)$
\item  $(\epsilon, \eta, q, r) = (\epsilon \neq 0, \eta \neq 0, q = 0, r \neq 0)$ such that
$(\alpha\delta + \beta\gamma){\eta} = {r} \xi (\alpha\gamma - \beta\delta)$
\end{itemize}

\noindent
(ii) $\alpha\beta - \gamma\delta \neq 0$ and $\alpha\gamma - \beta\delta = 0$

$(\epsilon, \eta, q, r) = (\epsilon\neq 0, \eta = 0, q = 0, r\neq 0)$
\vspace{1em}

\noindent
(iii) $\alpha\beta - \gamma\delta = 0$ and $\alpha\gamma - \beta\delta \neq 0$

$(\epsilon, \eta, q, r) = (\epsilon=0, \eta\neq 0, q\neq 0, r=0)$
\vspace{1em}

\noindent
(iv) $\alpha\beta - \gamma\delta = 0$ and $\alpha\gamma - \beta\delta = 0$

There are three possibilities:

(a) $(\alpha=0, \beta\neq 0, \gamma\neq 0, \delta=0)$

\ $(\epsilon, \eta, q, r)$ such that
$2\beta\gamma{\epsilon}{\eta} = {q}{r}$
\vspace{1em}

(b) $(\alpha\neq 0, \beta=0, \gamma=0, \delta\neq 0)$

\ $(\epsilon, \eta, q, r)$ such that
$-2\alpha\delta{\epsilon}{\eta} = {q}{r}$
\vspace{1em}

(c) $(\alpha=\delta \neq 0, \beta=\gamma\neq 0, \alpha\neq \beta)$

When $16 {\alpha^2}{\beta^2}{\xi^2} - 1 \neq 0$:

$(\epsilon, \eta, q, r)$ such that
$2(\alpha^2 - \beta^2){\epsilon}{\eta} = (16 {\alpha^2}{\beta^2}{\xi^2} - 1){q}{r}$

When $16 {\alpha^2}{\beta^2}{\xi^2} - 1 = 0$:

$(\epsilon, \eta, q, r)$ such that
${\epsilon}{\eta} = 0$

\subsection{\underline{$\alpha\delta - \beta\gamma \neq 0$ and $\xi = 0$}}
\vspace{1em}

\subsubsection{$\alpha\delta + \beta\gamma = 0$} \
\vspace{-1em}

No constraints on $(\epsilon, \eta, q, r)$ 

\subsubsection{$\alpha\delta + \beta\gamma \neq 0$} \
\vspace{-1em}

$(\epsilon, \eta, q, r)$ such that
$2(\alpha\delta - \beta\gamma){\epsilon}{\eta} = (\alpha\delta + \beta\gamma){q}{r}$

\subsection{\underline{$\alpha\delta - \beta\gamma = 0$ and $\xi = 0$}}
\vspace{1em}

\subsubsection{$\beta\gamma = 0$} \
\vspace{-1em}

No constraints on $(\epsilon, \eta, q, r)$ 

\subsubsection{$\beta\gamma \neq 0$} \
\vspace{-1em}

$(\epsilon, \eta, q, r)$ such that ${q}{r} = 0$

\subsection{\underline{$\alpha\delta - \beta\gamma = 0$ and $\xi \neq 0$}}
\vspace{1em}

\subsubsection{$\xi = 1$} \
\vspace{-1em}

No constraints on $(\epsilon, \eta, q, r)$ (This is the separable case.)

\subsubsection{$\xi \neq 1$} \
\vspace{-1em}

(i) $\beta\gamma - {\xi^2}(\alpha\gamma + \beta\delta)(\alpha\beta + \gamma\delta)=0$

No constraints on $(\epsilon, \eta, q, r)$

(ii) $\beta\gamma - {\xi^2}(\alpha\gamma + \beta\delta)(\alpha\beta + \gamma\delta)\neq 0$

$(\epsilon, \eta, q, r)$ such that ${q}{r} = 0$
\vspace{1em}

\noindent
Q.E.D.

\section{Example: Bell States}

A set of Bell state vectors is a representative example of entangled state vectors for the $2\times 2$ quantum system. In our notation, it consists of the following four state vectors.
\begin{equation}
\ket{\Phi^{\pm}}= {1\over\sqrt{2}}(\ket{a_1}\otimes\ket{b_1} \pm \ket{a_2}\otimes\ket{b_2})
\equiv {1\over\sqrt{2}}(\begin{bmatrix}
1 \\
0
\end{bmatrix}_A \otimes
\begin{bmatrix}
1 \\
0
\end{bmatrix}_B 
\pm 
\begin{bmatrix}
0 \\
1
\end{bmatrix}_A
\otimes
\begin{bmatrix}
0 \\
1
\end{bmatrix}_B), 
\label{bell1}
\end{equation}

\begin{equation}
\ket{\Psi^{\pm}}= {1\over\sqrt{2}}(\ket{a_1}\otimes\ket{b_2} \pm \ket{a_2}\otimes\ket{b_1})
\equiv {1\over\sqrt{2}}(\begin{bmatrix}
1 \\
0
\end{bmatrix}_A \otimes
\begin{bmatrix}
0 \\
1
\end{bmatrix}_B 
\pm 
\begin{bmatrix}
0 \\
1
\end{bmatrix}_A
\otimes
\begin{bmatrix}
1 \\
0
\end{bmatrix}_B). 
\label{bell2}
\end{equation}

When we apply our theorem to these state vectors, we obtain that the
following set of two observable operators leads to zero correlation.

\begin{itemize}
\item The Bell state vectors in Eq. (\ref{bell1}) correspond to the classification 
${\it{3.1.2}}$(iv-b): 

$-2\alpha\delta{\epsilon}{\eta} = {q}{r}$.

\item The Bell state vectors in Eq. (\ref{bell2}) correspond to the classification 
${\it{3.1.2}}$(iv-a): 

$2\beta\gamma{\epsilon}{\eta} = {q}{r}$.
\end{itemize}

For a very simple example, with ${q}={r}= m(\neq 0)$, we can have the following pairs of operators yielding zero correlations for the Bell states:

\begin{eqnarray}
For \{ \ket{\Phi^{+}}, \ket{\Psi^{-}} \}: \quad 
{\mathcal{Q}_A}=
\begin{bmatrix}
Q_0 - m & m \\
m & Q_0 + m
\end{bmatrix}
= Q_0 {\bf{1}} + m (\sigma_x - \sigma_z)\nonumber\\
{\mathcal{R}_B}= 
\begin{bmatrix}
R_0 + m & m \\
m & R_0 - m
\end{bmatrix} 
= R_0 {\bf{1}} + m (\sigma_x + \sigma_z)\nonumber
\end{eqnarray}

\begin{eqnarray}
For \{ \ket{\Phi^{-}}, \ket{\Psi^{+}} \}: \quad 
{\mathcal{Q}_A}=
\begin{bmatrix}
Q_0 + m & m \\
m & Q_0 - m
\end{bmatrix}
= Q_0 {\bf{1}} + m (\sigma_x + \sigma_z)\nonumber\\
{\mathcal{R}_B}=
\begin{bmatrix}
R_0 + m & m \\
m & R_0 - m
\end{bmatrix}
= R_0 {\bf{1}} + m (\sigma_x + \sigma_z)\nonumber
\end{eqnarray}
In the above, we have used the identity operator and the Pauli matricies
\begin{equation}
{\mathcal{\sigma}_x}=
\begin{bmatrix}
0 & 1 \\
1 & 0
\end{bmatrix}
, \quad
{\mathcal{\sigma}_z}=
\begin{bmatrix}
1 & 0 \\
0 & -1
\end{bmatrix}.\nonumber
\end{equation}

\section{Discussion}
\subsection{Quantum pigeonhole effect}

From a broader perspective, the theorem presented here is only one example of the intricate relations among quantum and classical probability concepts. The ``quantum pigeonhole effect'' recently proposed by Aharonov et. al. \cite{aharonov} sheds even more light on the deeper characteristics of quantum entanglements by
creating correlations from separable product states, which are normally considered as non-entangled. 

Classically the pigeonhole principle states that if we have a number of pigeons to be placed in a smaller number of holes, at least one hole must contain multiple pigeons. The analogous system is considered in quantum mechanics with three two-state quantum particles (pigeons), each in a superposition of two (hole) states (a quantum $3\times2$ system). Computing correlations with cleverly chosen different pre- and post-selected product states, surprisingly, shows that no pair of particles can be in the same quantum (hole) state. This means that the pigeonhole principle in some cases breaks down in quantum mechanics. It also shows that there are new aspects of quantum entanglement that are not apparent in product states. Experimental work with three single photons transmitted through two polarization channels indicates that this quantum pigeonhole effect is real \cite{chen}.

In a way, our theorem and the pigeonhole effect point in opposite directions: the former creates zero correlations from entangled states, while the latter shows the existence of correlations from separable states. Both are, however, illuminating the borders between classical and quantum systems.

\subsection{Other issues}

We would like to return to our theorem and discuss a couple of points.

We again note that our derivation of a zero--correlation condition is only one possibility, and other
choices are possible. For example, two options are derived in 
classification ${\it{3.1.2}}$(i). Exploring and categorizing  different types of solutions is left for the future.

In general, one does not have a priori knowledge of the quantum state vector, which makes the construction of
observable operators with zero-correlations difficult. On the other hand, if one can infer or conjecture the quantum state of a $2\times2$ system,
our construction scheme can help to design confirmation measurements.

We conjecture, by inference from classical probability, that the theorem holds for two-particle systems with higher-dimensional basis. We may also extend the theorem to more particle systems, and/or mixed quantum states. Proofs for such extensions are, however, yet to be explored.

Finally, it is well known in classical probability theories that statistical independence entails zero correlation,
but the converse is not true (e.g., \cite{feller}). In this sense, statistical independence is a tighter concept than zero correlation. 
For the $2\times2$ system, however, they are equivalent classically, and our result shows that
this ``classical common knowledge'' holds  rather in quantum theory: separability entails zero correlation but the
converse is not true. The $2\times2$ systems have been extensively studied  
 in investigating foundations of 
the quantum mechanics and quantum measurement theories (e.g., \cite{araki,yanase,popsecu,hess,allah}). It is interesting
that this familiar system provides our theorem that separates the classical and quantum theories in a peculiar manner. 

\section*{Acknowledgments}

The author would like to thank Philip M. Pearle, Professor Emeritus of Hamilton College, for his comments and encouragement.
This work was supported by funding from Ohagi Hospital, Hashimoto, Wakayama, Japan, and by Grant-in-Aid for Scientific Research from the Japan Society for the Promotion of Science Nos.19H01201 and 18H04443.

\section*{Appendix}

After the original version of this paper is placed in the arXiv, I received a comment from
Dr. Shuming Cheng that my result can be extended to the entangled mixed states of the $2\times2$ system. I outline his argument in the following.

Any density matrix for the $2\times2$ systems can be written using the Pauli matrices as follows.

\begin{equation}
\rho_{AB}={1\over 4}({\bf{1}_A}\otimes {\bf{1}_B}+{\vec{a}}\cdot{\vec{\mathcal{\sigma}}} \otimes {\bf{1}_B} + {\bf{1}_A}\otimes {\vec{b}}\cdot{\vec{\mathcal{\sigma}}} +\sum_{ij} F_{ij}{\mathcal{\sigma}}_i\otimes {\mathcal{\sigma}}_j)
\label{density}
\end{equation}
where ${\bf{1}}$ is the $2\times2$ identity matrix, ${\vec{a}},{\vec{b}}$ are vectors consists of 3 real numbers ($\cdot$ is the inner product),
and $F_{ij}$ are real number elements of a $3\times3$ matrix ${\mathcal{F}}$,
and ${\vec{\mathcal{\sigma}}} = ({\mathcal{\sigma}_x}, {\mathcal{\sigma}_y},{\mathcal{\sigma}_z})$ is a vector with the Pauli matrices.
\begin{equation}
{\mathcal{\sigma}_x}=
\begin{bmatrix}
0 & 1 \\
1 & 0
\end{bmatrix}
, \quad
{\mathcal{\sigma}_y}=
\begin{bmatrix}
0 & -i \\
i & 0
\end{bmatrix}, \quad
{\mathcal{\sigma}_z}=
\begin{bmatrix}
1 & 0 \\
0 & -1
\end{bmatrix}.\nonumber
\end{equation}

Now, the two quantum observable operators ${\mathcal{Q}_A}, {\mathcal{R}_B}$ can be also expressed using the Pauli matrices up to a scale as
\begin{equation}
{\mathcal{Q}_A} = {1\over 2}({\bf{1}_A} + {\vec{x}}\cdot{\vec{\mathcal{\sigma}}}), \quad {\mathcal{R}_B} = {1\over 2}({\bf{1}_B} + {\vec{y}}\cdot{\vec{\mathcal{\sigma}}})
\label{obs}
\end{equation}
where ${\vec{x}},{\vec{y}}$ are three dimensional real vectors.

Then, for operators $\mathcal{X} = {\mathcal{Q}_A}\otimes{\bf{1}}_B$ and $\mathcal{Y} = {\bf{1}}_A\otimes\mathcal{R}_B$, we can calculate the expectation values as follows;
\begin{equation}
\langle \mathcal{X}\mathcal{Y} \rangle = {\mathrm{Tr_{AB}}}[\rho_{AB}\mathcal{X}\mathcal{Y}]={1\over 4}(1+
{\vec{a}}\cdot{\vec{x}} + {\vec{b}}\cdot{\vec{y}} + {\vec{x}}\cdot{\mathcal{F}}\cdot{\vec{y}}),\nonumber
\end{equation}
and
\begin{equation}
\langle \mathcal{X} \rangle ={1\over 2}(1 + {\vec{a}}\cdot{\vec{x}}),\quad 
\langle \mathcal{Y} \rangle ={1\over 2}(1 + {\vec{b}}\cdot{\vec{y}}). \nonumber
\end{equation}
This leads to
\begin{equation}
\langle \mathcal{X}\mathcal{Y} \rangle - \langle \mathcal{X} \rangle\langle \mathcal{Y} \rangle = {1\over 4}({\vec{x}}\cdot{\mathcal{F}}\cdot{\vec{y}}+ ({\vec{a}}\cdot{\vec{x}})({\vec{b}}\cdot{\vec{y}})) = {1\over 4}{\vec{x}}\cdot({\mathcal{F}} - {\vec{a}}\cdot{{\vec{b}^{\hspace{0.05cm} \mathsf{T}}}})\cdot{\vec{y}},
\end{equation}
where ${\vec{a}}\cdot{{\vec{b}^{\hspace{0.05cm} \mathsf{T}}}}$ is the outer product of ${\vec{a}}, {{\vec{b}}}$. 
Hence, the zero correlation condition is given as 
\begin{equation}
{\vec{x}}\cdot({\mathcal{F}} - {\vec{a}}\cdot{{\vec{b}^{\hspace{0.05cm} \mathsf{T}}}})\cdot{\vec{y}} = 0.
\label{orth}
\end{equation}

Given any density matrix (\ref{density}), 
the corresponding real $3\times3$ matrix $({\mathcal{F}} - {\vec{a}}\cdot{{\vec{b}^{\hspace{0.05cm} \mathsf{T}}}})$ is fixed. 
One can always find a pair of $({\vec{x}},{\vec{y}})$, i.e., two observable operators in (\ref{obs}),
satisfying (\ref{orth}) as it is an orthogonality relation in the real three--dimensional space.



%

\end{document}